\documentclass[pre,showpacs]{revtex4}
\usepackage{graphics} 
\newcommand{\bea}{\begin{eqnarray}}
\newcommand{\eea}{\end{eqnarray}}
\newcommand{\bfv}{\bf v}
\newcommand{\N}{\nabla}
\newcommand{\cd}{\cdot}
\newcommand{\D}{\Delta}
\newcommand{\K}{\kappa}
\newcommand{\im}{\imath}

\begin{document}
\title{Lattice gas automaton approach \\
       to ``Turbulent Diffusion''}
\author{Jean Pierre Boon} 
\email{jpboon@ulb.ac.be}
\homepage{http://poseidon.ulb.ac.be/} 
\author{Eric Vanden Eijnden} 
\email{eve2@cims.nyu.edu}
\author{David Hanon}  
\email{david.hanon@ulb.ac.be}
\affiliation{{\em D\'epartement de Physique, C.P. 231} \\
{\em Universit\'e Libre de Bruxelles, 1050 Bruxelles, Belgium}}
\date{{\em CHAOS, SOLITONS, and FRACTALS}, {\bf 11}, 187-192 (2000)}

\begin{abstract}

A periodic Kolmogorov type flow is implemented in a lattice gas automaton.
For given aspect ratios of the automaton universe and within a range of 
Reynolds number values, the averaged flow evolves towards a stationary 
two-dimensional $ABC$ type flow. We show the analogy between the streamlines
of the flow in the automaton and the phase plane trajectories of a 
dynamical system. In practice flows are commonly studied by seeding the 
fluid with suspended particles which play the role of passive tracers.
Since an actual flow is time-dependent and has fluctuations, 
the tracers exhibit interesting intrinsic dynamics. When tracers are 
implemented in the automaton and their trajectories are followed, we find 
that the tracers displacements obey a diffusion law, with ``super-diffusion''
in the direction orthogonal to the direction of the initial forcing.


\end{abstract}

\maketitle

\section{Introduction}
\label{sec_intro}

In an incompressible fluid undergoing two-dimensional flow, the equations 
describing the trajectories of the fluid elements in terms of the stream 
function $\psi $ take the Hamiltonian form

\bea
\label{hamil}
\dot{x} \equiv v_x &=& - \partial_y \psi (x,y;t) \,, \nonumber \\
\dot{y} \equiv v_y &=&   \partial_x \psi (x,y;t) \,,
\eea 
where $\bf v$ is the fluid velocity.  In practice the
trajectories of the fluid elements are visualized by seeding the fluid with 
suspended particles which act as  passive tracers, whose velocity is given
by Eqs.(\ref{hamil}). If the flow were stationary, $\psi_{ss} = \psi (x,y)$, 
the system would be conservative and thus integrable, and the trajectories 
of the tracers would coincide with the curves of the contour plot of the 
stream function. However the actual flow possesses intrinsic fluctuations, 

\bea
\label{psi_fluc}
\psi (x,y;t) = \psi_{ss} (x,y) + \delta \psi (x,y;t)\,,
\eea
which in general precludes exact analytical solution of Eqs.(\ref{hamil}). 
A consequence of the presence of noise is that the tracers eventually
exhibit diffusive behavior. The problem that we address here is the 
evaluation of the tracer dynamics in the flow (\ref{psi_fluc}).

\section{Kolmogorov flow}
\label{sec_kolmogorov}

The Navier-Stokes equations for an incompressible fluid subject to an
external force $\bf F$ read

\bea
\label{nseq}
\partial_t {\bfv} +  ({\bfv} \cd {\N})\, {\bfv} & =& - \rho^{-1}\N P
           + \nu \N^2 {\bfv} + \bf F \,, \nonumber \\
{\N} \cd {\bfv} &=& 0 \,,
\eea
where $P$ is the hydrostatic pressure, $\rho$ the mass density, and $\nu$ 
the kinematic viscosity. When the flow is two-dimensional and $\bf F$ has 
the form

\bea
\label{force}
{\bf F} = {\Phi} {\bf 1}_x cos (\K y)\,, \;\;\;\;\;  \K = (2 \pi / L)\,,
\eea
where L is the extension of the system in the $y$-direction, Eqs.(\ref{nseq})
describe the periodic Kolmogorov flow \cite{zaslavsky}. 
Alternately, with $\psi$ defined by Eqs.(\ref{hamil}), 
the Navier-Stokes equations can be recast in the form

\bea
\label{psieq}
(\partial_t - Re^{-1} \N^2) \psi + J(\psi, \D\psi) = Re^{-1} cos (\K y)\,; 
\eea
here $Re = \psi_0 / \nu$ is the Reynolds number, where $\psi_0$ is defined
by  $\psi_{ss} = \psi_0 cos (\K y)$, and $J(\psi, \D\psi)$ is the Jacobian
\bea
\label{jacob}
J(\psi, \D\psi) = \partial_x \psi \partial_y \D\psi -
                  \partial_y \psi \partial_x \D\psi\,.
\eea

By linear stability analysis, one can show that there exists a critical
value of the Reynolds number where the system becomes unstable. This is
most easily seen by considering a perturbation to the velocity field

\bea
\label{perturb}
v_x =  v_x^{ss} + \delta v_x \,,\;\;\;   v_y = v_y^{ss} + \delta v_y  \,,
\eea
with
\bea
\label{vss}
v_x^{ss} = \frac {\Phi}{\nu \K^2}\; cos (\K y)\;\;\; , v_y^{ss} = 0\,,
\eea
and evaluating the perturbation evolution in response to the imposed forcing. 
We write the perturbation in terms of its spatial Fourier components as

\bea
\label{fourierx}
\delta v_x (x,y;t) = e^{\im q x} \phi (y;t) \,,\;\;\; 
\delta v_y (x,y;t) = e^{\im q x} \varphi(y;t) \,,
\eea
with 
\bea
\label{fouriery}
\phi (y;t) = \sum_{k=-\infty}^{\infty} e^{\im k y}\phi_k(t)\,,\;\;\;  
\varphi(y;t) = \sum_{k=-\infty}^{\infty} e^{\im k y}\varphi_k (t)\,.
\eea
We insert (\ref{perturb}) with (\ref{fourierx}) and (\ref{fouriery}) into 
the Navier-Stokes equations which are then Laplace-Fourier transformed to
obtain the characteristic equation and the dispersion equation; retaining 
only the modes $k=0,\,k=\K$, and $k=-\K$ (i.e. setting to zero the amplitudes 
of the  modes with $|k|> \K$), we obtain from the characteristic equation the
value of the critical Reynolds number
\bea
\label{reynolds} 
Re_c = {\sqrt 2} \,\, \nu \K^2 \frac{\K^2 + q^2}{(\K^2 - q^2)^{1/2}} \,,
\eea
and from the dispersion equation the modes
\bea
\label{modes}
\delta v_x (q,t) \sim \,e^{s_+t} \,,\;\; s_+ &=& -\nu \K^2 + 
\nu q^2\epsilon\,,
\\
\delta v_y (q,t) \sim  \,e^{s_-t} \,,\;\; s_- &=& -\nu q^2\epsilon\,,
\eea
with $\epsilon = 1 - (Re/Re_c)^2$. Equation (\ref{modes}) shows that the
mode $\delta v_y$ orthogonal to the direction of the forcing (along the 
$x$-axis) exhibits {\em critical slowing down}, i.e. $s_- \rightarrow 0$ when 
$Re \rightarrow Re_c$.  At $Re =  Re_c$, the Kolmogorov flow becomes unstable,
and beyond the bifurcation ($Re > Re_c$) goes into a 2-dimensional $ABC$
type flow \cite{zaslavsky2} with closed streamlines, separatrices, and 
infinite trajectories between separatrices, as illustrated in Fig.1.

\section{Lattice gas automaton dynamics}
\label{sec_lga}

The flow to be investigated is produced by a {\em lattice gas
automaton} \cite{lga} which we briefly describe.  A lattice gas
automaton can be viewed as a collection of particles residing in a 
discrete space, a regular $d$-dimensional lattice $\cal{L}$, where they move
at discrete time steps. Associated to each lattice node (with position 
denoted by ${\bf r}$) there is a finite set of channels (labeled by Latin 
indices $i,j,..$). Each of these channels corresponds to a discrete value of
the velocity $({\bf c}_i)$ that a particle positioned at the specified
node may have. An {\it exclusion principle} imposes that there be a maximum 
of one particle per channel. The  exclusion principle is important
because it allows a symbolic representation of the state of the system
in terms of bits, and of its dynamics in terms of operations over sets
of bits, which are easily implemented on a computer. The state of the
automaton at time $t$ is thus described by specifying the {\it configuration}
on each and every node, i.e. the set of bits $ \{n_i({\bf r},t) \}_{i=1}^{b}$
, $ {\bf r} \in {\cal L} $, for an automaton with $b$ channels per node.
The evolution of the automaton takes place in two stages : {\em propagation} 
and {\em collision}, applied sequentially at every time step. 
Particles are first moved according to 
their velocity: If channel $i$ at node ${\bf r} $ is occupied, the
propagation step displaces that particle to channel $i$ of node ${\bf
r} + {\bf c}_i \Delta t $. This updating is done synchronously throughout the
lattice. The second stage in the dynamics is a local collision step:
As a function of the pre-collisional configuration, a new
configuration is chosen by a prescription based on a set of (usually
stochastic) rules. This step is crucial
to determine the type of physics the automaton will exhibit at the
macro- and mesoscopic levels. In particular, the collision step should
preserve the quantities that are invariant under the dynamics of the
model system. For example, it is possible to construct thermal
automata whose rules are such that the number of particles, the
momentum and the energy remain unchanged by the collision step
\cite{GBL}. In  general, there will be several sets of collision rules 
consistent with the invariance of the specified constants of motion under 
the collision step. The choice of a particular set is then dictated by 
operational convenience, or by the need to explore a particular physical 
regime. Here we are interested in the
two-dimensional hydrodynamic regime and it suffices to implement a simple 
set of rules governing the automaton dynamics on a triangular lattice
\cite{FHP}. It can be shown, starting from the {\em microscopic} equations 
of the automaton, that, provided the symmetry of the lattice is 
sufficient, the {\em macroscopic} dynamics of the automaton is consistent
with the Navier-Stokes equations \cite{suarez}. It is one of the virtues
of the lattice gas automaton that on the sole basis of microscopic rules
in accordance with local invariance and symmetry properties, its macroscopic
behavior produces correct hydrodynamics.

Furthermore the lattice gas automaton exhibits two important features:
(i) It possesses a large number of degrees of freedom;  
(ii) Its Boolean microscopic nature combined with stochastic micro-dynamics
results in {\em intrinsic spontaneous fluctuations}, and it has been shown 
that these fluctuations capture the essentials of actual fluctuations in real 
fluids \cite{grosfils}. Therefore the lattice gas automaton can be considered
as a {\em reservoir of excitations} extending over a wide range of frequencies
and wavelengths.

\section{Tracer dynamics}
\label{sec_tracer}

Tracers are implemented as suspended particles in the following way. The
tracer, which we denote as a $t$-particle, is subjected to the cooperative 
effects of the fluid particles, and its dynamics results from the {\em local 
dynamics} of the automaton, that is from the combined effects of 
{\em deterministic} advection (due to the non-zero average velocity field 
resulting from the constraint imposed to the lattice gas) and {\em random} 
advection 
(due to the automaton intrinsic fluctuations). The $t$-particle undergoes
displacements according to the average velocity of the fluid particles 
computed over a local domain of $\cal{L}$ and over a number ($\beta $) 
of time steps of the automaton ($\beta \Delta t$).
In the absence of external force ($Re = 0$, that is 
$\psi_0 = 0$ in the notation of Section \ref{sec_kolmogorov}), the 
dynamics of the $t$-particle is governed solely by the automaton noise 
and the tracers undergo random motion leading to diffusive behavior over
long distances and long times, i.e. $\langle \delta x^2 (t) \rangle  = 
\langle \delta y^2  (t) \rangle = 2 D_0 t$, for $t \gg \Delta t$, where
$D_0$ will be referred to as the coefficient of {\em molecular diffusion},
and $\Delta t$ is the elementary time step of the automaton. The value of 
$\beta$ can be varied to modify the effect of noise and thereby to tune the 
coefficient of molecular diffusion $D_0$ of the $t$-particle. So the dynamics 
of the tracers can be represented in a space defined on a plane parallel to 
the $\cal{L}$-plane and on a time scale set by a time increment equal to
$\beta \Delta t$. 

We are now interested in the dynamics of the $t$-particles in a fluid subject
to an external force. Therefore we impose a bias to the velocity field of the
automaton by preparing its initial state by distributing the velocities of 
the fluid particles on each node of the lattice so that on the average we
obtain a periodic velocity profile according to Eqs.(\ref{vss}).
With periodic boundary conditions imposed on the automaton universe, the 
shear triggered by the velocity bias produces a stationary flow in the form of
a periodic array of vortices; at a moderately high value of the Reynolds 
number ($Re / Re_c \sim 3$) the  spatial structure of the streamlines 
becomes analogous to the topology of the two-dimensional $ABC$ flow. In the 
absence of noise, the $t$-particles would follow exactly the
streamlines, and their motion would be ballistic. However because 
intrinsic fluctuations are always present in the automaton (as they are in 
real fluids), the tracers dynamics can be strongly perturbed, and it is
only in the averaged automaton flow that their trajectories reflect the 
topology of the $ABC$ flow, as illustrated in Fig.2. 

When we consider the  large-scale, long-time limit ($t \gg L_*/ V_*$, where 
$L_*$ and $V_*$ are the vortex characteristic quantities), we can adopt a 
Fokker-Planck formulation for the $t$-particle distribution function 
$\langle f \rangle$, where the average is taken over the ensemble
of realizations and, by averaging over a sufficiently large region of space 
({\em homogenization} hypothesis \cite{homo}), we  obtain a  diffusion-type 
equation 
\cite{details} 

\bea
\label{diff_eq}
\partial_t \langle \tilde {f} \rangle\, =\, D^*_{\alpha \beta} 
\partial_\alpha \, \partial_\beta \langle \tilde {f}  \rangle \,,
\eea
where ${\bf D}^*= {\bf F}(V_*, L_*, D_0)$ is the {\em effective diffusion} 
coefficient of the $t$-particles in the flow. On the basis of a power law 
assumption, we infer by dimensional analysis  that \cite{crisanti}

\bea
\label{dimensional}
[D^*]\,=\, [D_0]^{\mu}\,[L_*]^{1-\mu}\,[V_*]^{1-\mu}\,,
\eea
indicating that for $|\mu| < 1$, and $D_0$ sufficiently small ($D_0 <
L_* V_*$, which is the case in the lattice gas automaton),
the elements $D^*_{\alpha \alpha}$ 
($D^*_{\alpha \beta}\,=\,D^*_{\alpha \alpha} \delta_{\alpha \beta}$)
should be larger than $D_0$, with an important quantitative difference
between $D^*_{\alpha \alpha}$ with $\mu > 0$, and $D^*_{\beta \beta}$
with $\mu < 0$.

We performed lattice gas automaton simulations as described above, and 
from measurements of the mean-square displacements of the $t$-particles   
 $\langle \delta x^2 (t) \rangle$ and $\langle \delta y^2  (t) \rangle $,
we find that over sufficiently long times they exhibit diffusive behavior,
as shown in Fig.3. Notice that in the absence of external force, the 
mean-square displacement of the $t$-particle (due to the sole effect of
the automaton noise) is isotropic and corresponds to  molecular diffusion 
(full line in  Fig.3).

The diffusion coefficients 
\bea
\label{diff_coef}
D^*_{\|} \equiv  D_{xx} & = &  \lim_{t \rightarrow \infty} 
\frac{\langle \delta x^2 (t) \rangle}{2 t} \;,\nonumber \\
D^*_{\bot} \equiv  D_{yy} &  = &  \lim_{t \rightarrow \infty} 
\frac{\langle \delta y^2 (t) \rangle}{2 t} \;.
\eea
can be evaluated quantitatively from the explicit 
expression of the flow and assuming a Gaussian form for the space and
time dependence of the second moment of the noise, which is compatible with 
the spectrum of the automaton fluctuations. We obtain \cite{details}

\bea
\label{diff_x_y}
D^*_{\bot} &=& \frac{D_0}{1 + \chi}\,,\;\;\;\chi = \frac{\bar{v}_x}
{{(\langle \delta v_x^2 \rangle)}^{\frac{1}{2}}}\,,\nonumber \\
D^*_{\|} &=& \frac{{\bar{v}_x}^2}{2 \kappa}\, \frac{1}{D^*_{\bot}}\,
\frac{1 + \frac{1}{2} \chi}{1 + \chi^2} \,+\,  D^*_{\bot}\,,
\eea
where $\langle \delta v_x^2 \rangle$ denotes the automaton fluctuations,
and $\bar{v}_x$ the amplitude of the forcing in the velocity field acting
on the $t$-particle: $\dot{x}(t) = \delta v_x(t) + \bar{v}_x cos (\kappa y)$.
In the limit $\bar{v}_x = 0$, one has $ D^*_{\bot} = D^*_{\|} = D_0$, and 
for $\bar{v}_x \neq 0 $, it follows from (\ref{diff_x_y}), that 
$D^*_{\bot} > D^*_{\|} > D_0$. We find qualititative agreement between 
these predictions and our simulation data (see Fig.3).

\section{Concluding comments}
\label{comments}

We have presented an automaton approach to the problem of ``turbulent
diffusion'' in a time-dependent flow with non-trivial average. We have shown 
the analogy between the stream lines in the averaged flow of the 
automaton and the phase plane trajectories of the corresponding dynamical 
system. For the full flow, we obtain agreement between the theoretical value 
of the diffusion coefficient of the tracers and the corresponding value 
computed from the lattice gas simulation data. The most important results are:
(i) above the critical value of the Reynolds number ($Re \simeq 3 Re_c$), 
the  tracers dynamics remains diffusive, and 
(ii) ``super-diffusion'' is observed in the direction orthogonal to the
direction of the forcing ($D^*_{\bot} \gg D_0$). Higher Reynolds number 
regimes and flows with non-stationary average are now being investigated.

\newpage

\begin{acknowledgments}

JPB acknowledges support by the {\em Fonds National de la Recherche 
Scientifique} (FNRS, Belgium). EVE was supported by the {\em Association
Euratom -- Etat Belge} (Belgium). DH has benefited from a grant from the 
{\em Fonds pour la Formation \`a la Recherche dans l'Industrie et 
l'Agriculture} (FRIA, Belgium).

\end{acknowledgments}

\begin{figure}
\resizebox{\textwidth}{!}{\includegraphics{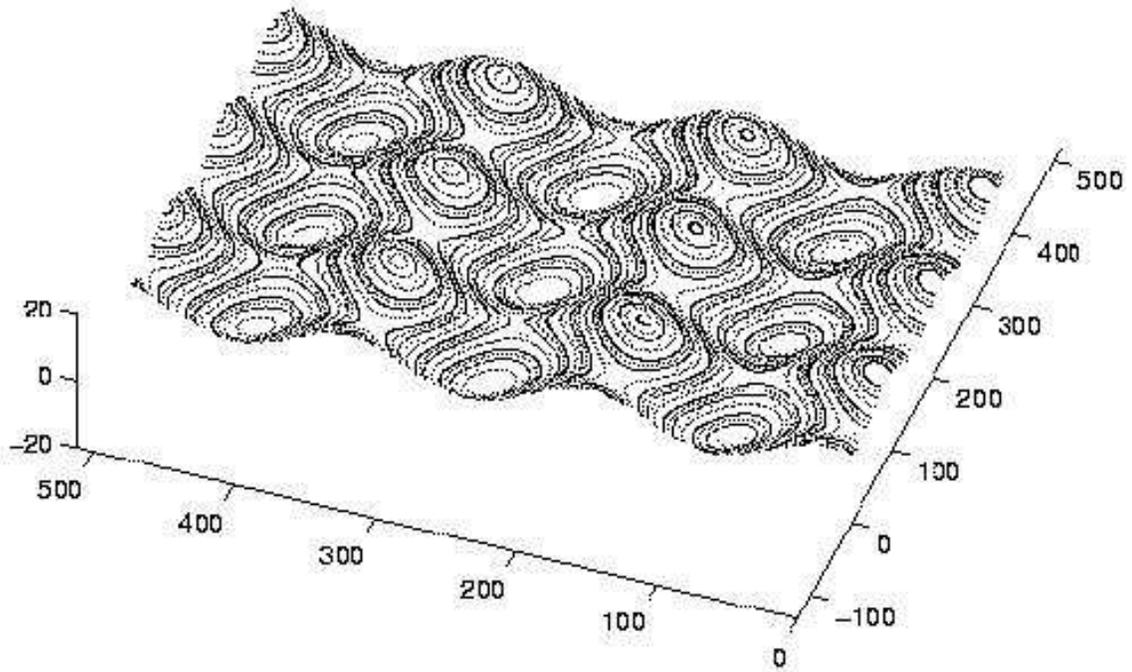}}
\caption{Contour plot of the stream function $\psi (x,y)$ at $Re/Re_c = 2.5$.
The streamlines shown here are obtained from the lattice gas automaton 
described in Section \ref{sec_lga} and subject to a periodic forcing
according to Eqs.(\ref{vss}). The horizontal plane is the $\cal{L}$-plane
(dimensions are given in lattice units) and the vertical axis gives the 
amplitude of the stream function $\psi (x,y)$ (in arbitrary units).}
\label{fig.1}
\end{figure}

\begin{figure}
\resizebox{\textwidth}{!}{\includegraphics{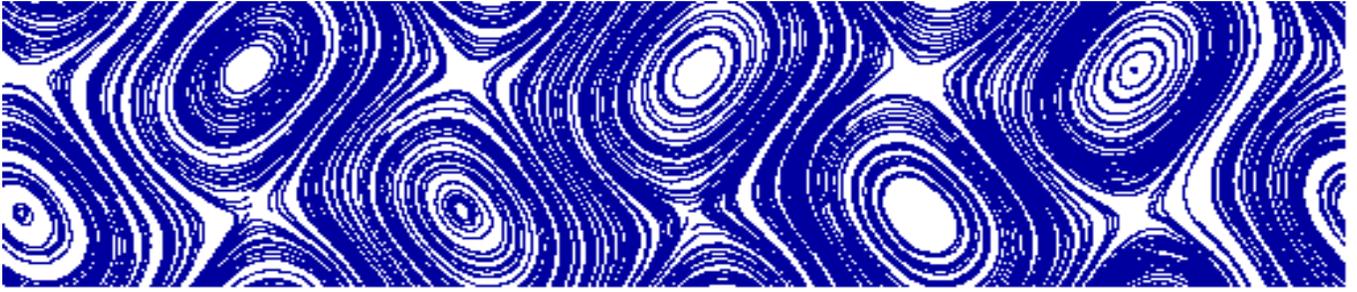}}
\caption{Tracer trajectories in the flow produced by the automaton with the
stream function shown in Fig.1. The tracer displacements have been
averaged over $1000$ time steps. Lattice size: $512 \times 128$ nodes;
average particle density: $0.2$  per channel; 
Reynolds number $Re = 2.5\; Re_c\,$; Mach number $Ma < 0.2$.}
\label{fig.2}
\end{figure}

\begin{figure}
\resizebox{\textwidth}{!}{\includegraphics{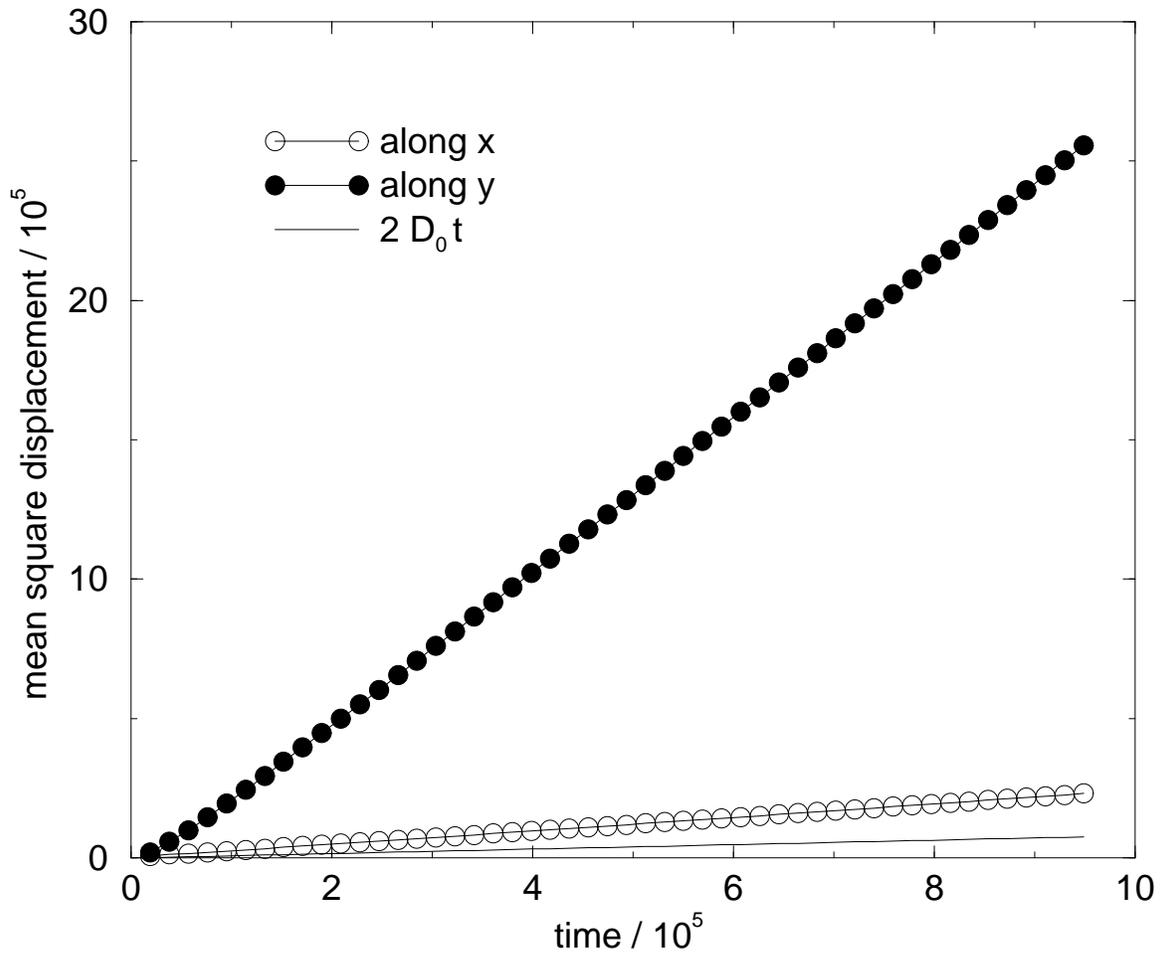}}
\caption{ Mean-square displacement of tracers measured as a function of
time. The data are the projections of the displacements along the $x$-axis 
($\circ $) and $y$-axis ($\bullet $), i.e. respectively along and 
perpendicular to the direction of the forcing 
($ {\bf F} = F_x(y) {\bf 1}_x $). The lower full line shows the diffusive 
behavior generated by the automaton fluctuations in the absence of external 
force. $\beta = 1$.
Distances are given in lattice units and time in automaton time steps.}
\label{fig.3}
\end{figure}


\begin{thebibliography}{99}
\bibitem{details} A complete analysis will be presented elsewhere.
\end{thebibliography}
\end{document}